\documentclass[12pt,epsf]{article}

\usepackage[dvips]{graphicx}

\newcommand{\beq}{\begin{equation}}
\newcommand{\eeq}{\end{equation}}
\newcommand{\bea}{\begin{eqnarray}}
\newcommand{\eea}{\end{eqnarray}}

\begin{document}
\thispagestyle{empty}
\vspace*{-15mm}
%----------
\baselineskip 1pt
\begin{flushright}
\begin{tabular}{l}
{\bf OCHA-PP-258}\\
{\bf February 2006}\\
{\bf hep-ph/0602203}
\end{tabular}
\end{flushright}
\baselineskip 24pt
\vglue 10mm

%%%%%%%%%%%%%%%%%%%%%%%%%%%%%%%%%%%%%%%%%%%%%%
%                Title
%%%%%%%%%%%%%%%%%%%%%%%%%%%%%%%%%%%%%%%%%%%%%%
\begin{center}
{\LARGE\bf
Meson Strings and Flavor Branes  
}
\vspace{7mm}

\baselineskip 18pt
{\bf Masako BANDO$^1$, Akio SUGAMOTO$^{2,3}$ and Sachiko TERUNUMA$^3$}
\vspace{2mm}

{\it
   $^1$Physics Division, Aichi University, Aichi 470-0296, Japan\\ 
$^2$Department of Physics, Ochanomizu University, Tokyo 112-8610, Japan\\
$^3$Graduate School of Humanities and Sciences, Ochanomizu University, Tokyo 112-8610, Japan
}\\
\vspace{10mm}
\end{center}
%%%%%%%%%%%%%%%%%%%%%%%%%%%%%%%%%%%%%%%%
%%%%%                              %%%%%
%%%%%          Abstract            %%%%%
%%%%%                              %%%%%
%%%%%%%%%%%%%%%%%%%%%%%%%%%%%%%%%%%%%%%%
\begin{center}
{\bf Abstract}\\[7mm]
\begin{minipage}{14cm}
\baselineskip 16pt
\noindent
%%%%%---------------------------------

We investigate the shape of meson strings in the five-dimensional 
curved space and the potential between the quark and anti-quark 
in a QCD-like string model based on D6 flavor branes in the presence of D4 color branes 
wrapping one of the compactified 
dimension on an $S^1$. 
  The flavor 
branes on which both ends of a meson string lie are assumed to be separated 
in this five dimensional space, depending on the values of the constituent quark masses.  It is 
shown in this picture that a 
 meson string with different flavors at two ends changes  shape 
at a critical distance.  There is, however,  no critical distance for a
 meson with the same flavor.  
At this critical distance, 
the potential between a quark and anti-quark with different flavors 
gives a point of reflection and  
changes shape near this point.  Accordingly, the attractive force between a 
quark and an anti-quark seems to 
become stronger when the distance between the flavor branes connecting 
meson strings becomes larger. 
This indicates that quark systems with different flavors can 
form high-density states. 

%%----------------------------------
\end{minipage}
\end{center}

%%%----------------------------------
%%%
%%%
%%%   Main body of the paper
%%%
%%%
%%%----------------------------------
\newpage
\baselineskip 18pt
\def\thefootnote{\fnsymbol{footnote}}
\setcounter{footnote}{0}
%%%%%%%%%%%
%%%Citation of reference is ~\cite{reference} .
%%%%Numbering of equation is \label{equation}, 
%%%%%%and referring the equation is(\ref{equation}) .
%%%%%%%%%%%%%

%%%--------------------------------------------
\section{Introduction}
The origin of families is one of the most important subjects 
of particle physics. 
It is well known that quarks and leptons appear as three repetitive families, 
each of which forms a multiplet of the strong, weak and 
electromagnetic gauge symmetries. 
The coupling constants of these gauge interactions are uniquely determined by the symmetry 
principle and are independent of the flavors. 
By contrast, flavor interactions determined by the Yukawa coupling to Higgs fields are 
mysterious; their coupling strength exhibits hierarchical structure depending on the flavor 
of the matter 
fields. 
 Many proposals for the origin of flavor have been made: 
It may originate from some family quantum number, Abelian or non-Abelian, by imposing a horizontal 
symmetry.  However, such hierarchical structure 
could not be derived from such symmetries. 
One promising idea is 
to extend our 4-dimensional space to a higher-dimensional space  
and to attribute the origin of flavor to 
the structure of the extra dimesions. 
If we can gain some information from such an idea with higher-dimensional space-time, we may be 
able to take an 
important step toward  the understanding of flavor. 

In the previous papers \cite{BKST, S}, we studied QCD strings based on the picture that a quark is 
identified as a colored flavored string, one end of which has a color  (color end) and the other end 
 of which has a flavor (flavor end). Therefore, the string of a free quark emerges from a point on 
a color brane and terminates at a point on a flavor brane.  Furthermore,
  apart from the stack of color branes, the flavor branes are   
considered to be separated flavor by flavor in the 
direction 
of the extra dimension.  Hereafter, we denote the coordinate of this extra dimension by $u$.  
A meson consists of two strings from a quark and an anti-quark, but they are joined to form a single 
meson string, because color ends of two strings can be annhilated,  and only flavor ends appear on 
individual flavor branes.  Similarly,  a baryon consists of three strings, with the flavor ends 
of the three strings located on individual flavor branes, 
and the color ends of the three strings, if the colors are 
red, green and blue, can annhilate to form a junction.  
In this way,  we studied the exotic 
hadron of pentaquarks in the previous papers~\cite{BKST, S}.
In such a picture, the hierarchical structure of the masses of quarks 
can be naturally explained by taking  proper allocations of 
flavor branes in the 
curved space. In addition,  the existence of a horizon-like 
singularity coming from the stack of $N_c$ color 
branes can account for  the non-perturbative linear potential 
bertween a quark and an anti-quark.   

In this paper, we study the shape of meson strings in the five-dimensional curved 
space and calculate the potential between the quark and anti-quark in detail. 
Leaving the details to Ref. \cite{BKST},  
we briefly explain the picture of flavor branes exising  
in the space extended to extra dimensions.  
Here we adopt a ealistic QCD-like model in the string theory developed by Witten and others 
~\cite{N=0a}.  In this model,  all the supersymmetries
are broken by the compactification of one of the extra dimensions on 
an $S^1$ which is wrapped by D4 color branes. 
In the model, the gauge theory is represented by open strings, while the gravity theory is represented by 
closed strings.  Because the two theories are dual, 
we can replace the non-perturbative QCD by the 
corresponding classical gravity theory.  
The original space of the gravity theory is flat. 
However, once heavy sheets of $N_c$ color ($D_4$) branes 
are placed perpendicular to the extra dimension $u$, 
it is 
deformed in this extra dimension.  
Because each endpoint of a meson string lies on a flavor brane, meson strings posseses 
different shapes, according to the variety of the flavors of the quarks on two ends.

Let us simplify the ten-dimensional space-time of string theory to a 
5-dimensional space-time, 
including only the extra coordinate $u$ 
in addition to the usual Minkowski space, $(t, z, {\bf x}_{\perp})$.  
The world volumes of the stack of
 $N_{c}$ color branes and the separately allocated flavor branes extend along the Minkowski space 
but are perpendicular to the $u$ direction.  
This curved space can be expressed as 
\begin{equation}
ds^2=f(u)(-dt^2+dz^2+d{\bf x}_{\perp}^2)+g(u)du^2,
\end{equation}
with the functions  $f(u)$ and $g(u)$. 
In a realistic QCD-like model, they are ~\cite{N=0a} 
\begin{eqnarray}
f(u)&=&(u/R')^{3/2},~~g(u)=(f(u)h(u))^{-1}, \nonumber \\
h(u)&=&1-(U_{KK}/u)^3,   \label{QCD like model}
\end{eqnarray}
with $R'$ and $U_{KK}$ given by
\begin{equation}
R'^{3}=2\pi\alpha_{c}N_{c}\alpha'/M_{KK},~~\mbox{and}~~ U_{KK}=\frac{8\pi}{9}\alpha_{c}N_{c}\alpha'M_{KK},
\end{equation}
where $(2\pi \alpha')^{-1}$ is the string tension of the starting Lagrangian, being $O(M_{\rm{Planck}}^2)$, 
and $\alpha_{c}$ is the QCD (with 
$N_c$ colors) coupling.  
The warp factor $f(u)$ is a monotonically increasing function of $u$.
\footnote{For reference, we also give the metric of the proto-type model of 
Mardacena~\cite{M}, possessing $N=4$ supersymmetries, 
\begin{equation}
f(u)=g(u)^{-1}=(u/R)^2, \label{maldacenametric}
\end{equation}
with the radius $\mathbf{R}$ of $AdS_5$ space  given by
\begin{equation}
R^4=8\pi\alpha_{c} N_{c} \alpha'^2,  \label{R^4}
\end{equation} 
In this case  $f(u)$ is also an increasing function of 
$u$. } 
In the following, we write $\alpha'$ (of the order of the Planck length squared) 
dependence explicitly. This was omitted in Ref. \cite{BKST}. 

In such a deformed space including the one extra dimension $u$, 
there exist  ``flavor branes" on  which the endpoints of the hadron strings 
 are placed, while quarks are 
expressed as open strings having one endpoint on the flavor brane and the other endpoint 
on the ``color brane''.
The shape of a hadron string chooses its most economical path in this curved space.  
It is analogous to a catenary 
under the influence of gravity on the earth.  
In our case, gravity is along the $u$-direction, 
and its nature is different from that of gravity on the earth.

Here we concentrate our attention on a meson string with 
different flavors at either end  and calculate their shapes and the potentials between the 
quark and anti-quark, especaily those with different flavors. 

\section{QCD strings in curved space}
First, we give a brief survey of the formulae  needed in this paper\cite{BKST}. 
The string action in the background curved space is expressed as  
\begin{equation}
S= \frac{1}{2\pi\alpha'} \int d\tau d\sigma 
\sqrt{-(\dot{X}^{M}\dot{X}_{M})(X'^{N}X'_{N})+(\dot{X}^{M}X'_{M})^2}, \label{action}
\end{equation}
where the dot and the prime represent derivatives  with respect to $\tau$ and $\sigma$, 
respectively, and $X^{M}(\tau,\sigma)$ 
describes the configuration of the worldsheet of a string with the two 
parameters  $\tau$ and $\sigma$.  Here, the outside space ({\it i.e.} the target space) of $X^{M}$ is curved, and the contraction $\dot{X}^{M}\dot{X}_{M}$, {\it etc.}, represents that with the metric $G_{MN}(x)\dot{X}^{M}\dot{X}^{N}$, ${\it etc}$.  

If we fix the parameterization of the worldsheet, 
choosing $\sigma=z$ and $\tau=t$, and take the static limit, $\dot{X}^{M}=0\, 
(M\ne 0)$, the string action (\ref{action}) in the time interval $\Delta t$ reads 
\begin{equation}
S=\frac{\Delta t}{2\pi\alpha'}\int dz L, 
\end{equation}
with the ``Lagrangian"
\begin{equation}
L=\sqrt{f(u)^2(1+({\bf x}'_{\perp})^2))+f(u)g(u)(u')^2}. \label{L}
\end{equation}

By regarding  $z$ as ``time", 
the following equation of motion is derived:  
\begin{equation}
u'=\frac{1}{(-H)} \sqrt{\frac{f}{g}\left( f^2-(-H)^2-({\bf p}_{\perp})^2 \right)},  
\label{eq:shape}
\end{equation}
with the three conserved quantities 
\begin{equation}
(-H)=\frac{f(u)^2}{L}, ~~\mbox{and}~~ ({\bf p}_{\perp})={\bf x}'_{\perp} (-H).   \label{pperp}
\end{equation}

The shape of the string $u(z)$ is determined by 
Eq.~(\ref{eq:shape}), once we fix the boundary conditions, and  
the coordinates of the endpoints of the string are denoted by $(U_1,Z_1)$ and $(U_2, Z_2)$.

The ``action"  during the change of $u$ from ${U_1}$ to ${U_2}$ is given by 
\begin{equation}
\frac{1}{2\pi\alpha'}\int_{U_1}^{U_2} dz L =\frac{1}{2\pi\alpha'}\int_{U_1}^{U_2} du \frac{1}{u'} \frac{f(u)^2}{(-H)}, 
\label{actionunittime}
\end{equation}
and thus the energy stored inside a string is  
\begin{equation}
E=\frac{1}{2\pi\alpha'} \int^{U_2}_{U_1}du \sqrt{ 
\frac{f(u)^3g(u)}{f(u)^2-(-H)^2-({\bf p}_{\perp})^2}}. \label{Energy}
\end{equation}
We can also estimate the change of $z$ from ${U_1}$ to ${U_2}$ as
\begin{equation}
r=Z_2-Z_1=\int^{U_2}_{U_1}~\frac{du}{u'}=(-H) \int^{U_2}_{U_1}du~~\sqrt{ 
\frac{g(u)}{f(u) \left(f(u)^2-(-H)^2-({\bf p}_{\perp})^2 \right)}},  \label{z}
\end{equation}
with the string shape determined by Eq.~(\ref{eq:shape}).
Note that this general formulation is also applicable to 
the more complicated web-like exotic hadrons and string systems.

\section{Meson string}
Now let us consider a meson string representing a bound state of $(q_1\bar q_2)$.  
This is a string connecting the 
flavor brane at ($u=U_1,\ z=Z_1)$ of quark $q_1$ and the flavor 
brane at ($u=U_2,\ z=Z_2)$ of anti-quark $\bar q_2$. 
We take $z$ to be the direction in which the string is streched.  In our picture, 
 quarks are not point particles  but also strings. 
However, it is acceptable to regard point particles $q_1$ and $\bar q_2$ as being  
placed on the flavor branes at $u=U_1$ and $u=U_2$, respectively, with 
the gluonic QCD string connecting them.  
Then, our meson string can be understood as the entire system consisting of the quark, anti-quark and the gluonic QCD string conecting them. Then, the distance $r=Z_2-Z_1$ can be interpreted 
as the distance between the point quark $q_1$ and the anti-quark $\bar q_2$ 
in the ordinary Minkowski space. 
We can choose  ${\bf x}_{\perp}={\bf 0}$ for mesons. 
Then, the string connecting the quark $q_1$ and
 the anti-quark 
$\bar q_2$ appears on the $(z, u)$-plane. 
Then, the equation of motion takes the simple form  
\begin{equation}
u'\equiv \frac{du}{dz}=\frac{1}{f_0^2} \, 
\sqrt{\frac{f(u)}{g(u)}\left( f^2(u)-f_0^2 \right)},  
\label{eq:mesonshape}
\end{equation}
with 
\begin{equation}
E=\frac{1}{2\pi\alpha'}\, \int^{U_2}_{U_1}du \sqrt{ 
\frac{f(u)^3g(u)}{f(u)^2-f_0^2}}. \label{mesonEnergy}
\end{equation}

Here, we fix the constant value $(-H)$ at the point $u=U_0$; that is, we have 
$(-H)=f(U_0)\equiv f_0$. 
This means that $u=U_0$ is a stationary point of $u$ giving $u'=0$.  We can always find $Z_0$ for 
which 
$u(Z_0)=U_0$ and $u'(Z_0)=0$ hold, and therfore 
we can take $Z_0=0$ hereafter without loss of generality. 

There are two cases here. The first is that in which  
the stationary point, $(z, u)=(0, U_0)$, giving $u'=0$, exists on 
the string streching between 
their ends.  The second is that in which  
the stationary point $(0, U_0)$ is  not on the string but is outside 
the string. 
In the former case, if $Z_1 \le 0$, then $Z_2 \ge 0$, or vice versa. In the latter case, 
 both $Z_1$ 
and $Z_2$ are positive or negative.  In order to find a string profile, it is useful to note 
that the solution of Eq.~(\ref{eq:shape}) is 
symmetric under the exchange $z\rightarrow -z$, so that 
if we find a solution $(z, u=u(z))$, then $(-z, u=u(z))$ is also a solution.  Therefore, we 
have only to solve the equation for $z \le 0$. In Eq.(\ref{eq:mesonshape}), we have 
$u'\ge 0$, and thus 
if we start to solve the solution from the point $(z=0, U_0)$, 
$u(z)$ increases monotonically in the positive $z$ direction 
($0 \le z \le Z_2)$.    
We have to prepare two such solutions with $0 \le z \le Z_1$ and 
$0 \le z \le Z_2$. 
 In the case that there is a stationary 
point on a meson string, we obtain a solution of the string profile 
by connecting at $z=0$ the above two solutions, the solution with 
$0 \le z \le Z_2$, and the reversed solution of $0 \le z \le Z_1$ 
 with respect to the $u$-axis  (by exchanging $z$ and $-z$). 
 In the 
other case, in which the stationary point lies outside the meson string, 
the profile of the meson string is the $U_1 \le u \le U_2$ 
part of the solution of $0 \le z \le Z_2$, 
where we have assumed $U_1 \le U_2$.   
The former string is  a  combination of the string from 
$(-|Z_1|, -|U_1|)$ 
to $(0,U_0)$ and that from $(0,U_0)$ to  $(Z_2, U_2)$. 
On the other hand,  in the latter case, the string can be expressed 
as that connecting the points  $(0,U_0)$ and  $(Z_2, U_2)$ 
by substructing the string from  $(0,U_0)$ to  $(Z_1, U_1)$. 
Here,  the minimum value $U_0$ is determined as a function of their 
separation parameter,  $r=Z_2-Z_1$,  via Eq.~(\ref{z}). 

Now, from Eq.~(\ref{z}),  the separation parameter reads 
\begin{eqnarray}
r=Z_2-Z_1=f_0 \left[ \int_{U_0}^{U_2} \pm \int_{U_0}^{U_1}\right] du
\sqrt{\frac{g(u)}{f(u) \left(f(u)^2-f_0^2 \right)}},
\label{stringlength}
\end{eqnarray}
and the energy can be written from (\ref{Energy}) as  
\begin{eqnarray}
E=\frac{1}{2\pi\alpha'}\, \left[ \int_{U_0}^{U_2} \pm \int_{U_0}^{U_1}\right]
du \sqrt{ \frac{f(u)^3g(u)}{f(u)^2-f_0^2}}.
\label{totenergystring}
\end{eqnarray}
In the above expressions, we take the signature $+$ in the first case, $Z_1\le 0 \le Z_2$, 
 and 
$-$ in the second case,  $0 \le Z_1 \le Z_2$. 

The string shape is determined in such a manner as to minimize this 
energy for a given separation $r$ between a quark and an anti-quark. 
\begin{figure}
\begin{center}
\includegraphics[width=15cm,clip]{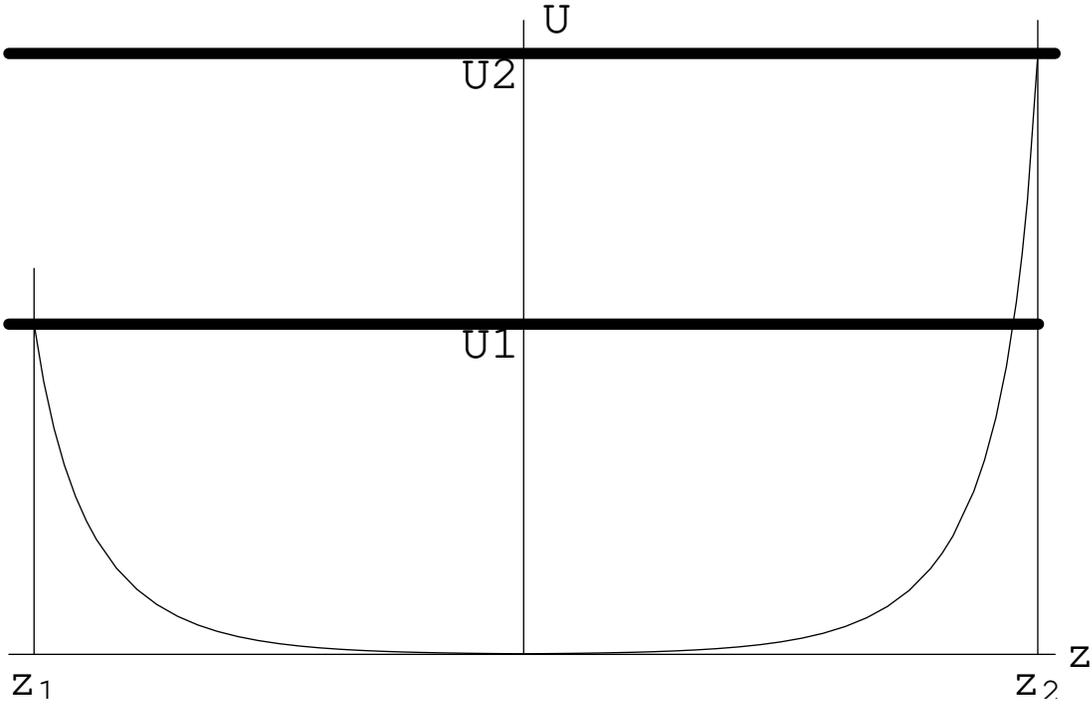}
\end{center}
\caption{Typical profile of the shape of a
 string connecting $q_{1}$ and $\bar q_{2}$  in the QCD-like model.}
\label{mesonstring}
\end{figure}
%---------------
A typical shape of a string for the case of a relatively large separation $r$ is depicted 
in Fig. 1, where 
the endpoints of the string terminate on the first brane and the second brane at 
$(Z_1,U_1)$ and $(Z_2,U_2)$, respectively. 

In the large separation limit($r\to \infty$), therefore, the 
string departing from the quark at $u=U_1$  quickly goes down vertically to the lowest possible value, $u=U_0\approx U_{\rm KK}$,  and 
then moves horizontally from $Z_1$ to $Z_2$,  
approximately maintaining the relation  $u \approx U_{\rm KK}$. 
Then it goes up to the anti-quark point at $u=U_2$.
This means that the basic shape of the meson string at a large separation is rectangular.

In order to understand this almost rectangular shape of the 
meson string, the following expression of the energy is useful: 
\begin{equation}
E= \frac{1}{2\pi\alpha'} \int^{U_2}_{U_1} \sqrt{ 
f(u) \left(f(u)dz^2+g(u)du^2\right)}.
\label{useful energy}  
\end{equation}
The approximately rectangular shape consists of two parts, the vertical and 
horizontal parts.  Using the above expression of energy,  we find 
\begin{equation}
E(\textrm{basic rectangular})\sim  E_{\rm{vertical}} + E_{\rm{horizontal}},
\end{equation}
where
\begin{equation}
E_{\rm vertical} 
=\frac{1}{2\pi\alpha'}\left[\int^{U_2}_{ U_{KK}}
+ \int^{U_1}_{ U_{KK}}\right]
\sqrt{f(u)g(u)}du ,
\end{equation}
\begin{equation}
E_{\rm horizontal} 
= \frac{1}{2\pi\alpha'} \int^{Z_2}_{Z_1} f(u)dz .
\label{horizontalgeneral}
\end{equation}
Because $f(u)$ is an increasing function of $u$, $f(u)$ is smallest at $u=U_{KK}$.  
Here, the region 
of $u$ is restricted by $u \ge U_{KK}$, 
due to the existence of the horizon-like singularity at $u=U_{KK}$.  
Therefore, the horizontal part on which 
$u$ is as small as possible near $u=U_{KK}$ is energetically favored. 

Let us define the quark mass (the constituent quark mass) $m_i$ as the vertical part of the string energy in the limit of large separation, 
namely the energy of the string stretching along the vertical line from the quark at $U_i$ on the 
$i$-th flavor 
brane to the horizon-like brane at $u=U_{\rm KK}$:   
\begin{equation}
m_i \equiv   \frac{1}{2\pi\alpha'} \int^{U_i}_{U_{KK}}\sqrt{f(u)g(u)}du\, . 
\label{quarkmass}
\end{equation}
Then, the remaining energy obtained after subtracting the quark masses can be understood as 
the potential energy $V(r)$ between the quark and anti-quark,  
\begin{equation}
V(r)= E_{tot}(r)-m_{q_1}-m_{q_2}\, .
\end{equation}
In the limit of large separation, the basic rectangular shape appears, and its horizontal part 
gives a typical potential energy, 
\begin{equation}
V(r) \approx \frac{1}{2\pi\alpha'} \int^{Z_2}_{Z_1} f(u)
\approx \frac{1}{2\pi\alpha'}\,\, {f(U_{KK})}\,  r,\quad  r=Z_2-Z_1, 
\label{horizontal}
\end{equation}
Therefore, in the limit of large separation, 
the QCD-like string model yields a linear potential proportional to $r$  with the 
coefficient\footnote{
Note that in  Maldacena's prototype model \cite{M}, there is no horizon-like singularity, 
namely  $f(U_{\rm MIN})=f(0)=0$.  Therefore, 
 in this case,  
the linear potential does not appear, 
and the potential becomes Coulomb-like, which reflects the conformal symmetry 
 at large distance.}  
\begin{equation}
k= \frac{ f(U_{KK})}{2\pi\alpha'}.
\label{QCDlikemodel}
\end{equation}

\section{Parameters of the QCD meson string}
Now let us determine the parameters appearing in the QCD-like string model from the 
information connecting the QCD strings and hadron phenomena. 
 
We denote the tension of QCD strings as $k$,  which is expressed in 
terms of the Regge slope, $\alpha'_{\rm{QCD}}$,  
\footnote{The parameter $\alpha' _{\rm{QCD}}$ 
in the meson potential is the Regge slope of QCD
strings, and it is different from the tension $\alpha'$ appearing 
in our original action. The latter is of the order of the Planck length squared.} 
as  
\begin{equation}
 k=\frac{1}{2\pi\, \alpha'_{\rm{QCD}}}.
\end{equation}
Let us derive this relation. First, if we rotate the QCD 
 string relativistically and classically estimate its energy 
(or its mass $M$) and its angular momentum $J$, 
then we have the relation coming from the so-called Regge trajectory, 
\begin{equation}
 J=\alpha'_{\rm{QCD}}\, M^2=\frac{1}{2\pi k} M^2.
\end{equation}
Next, utilizing the experimental data for the $J$ dependence of $M^2$ (Regge slope)~\cite{Reggeslopehara},we fix  $\alpha'_{\rm{QCD}}=0.9 \rm{(GeV)^{-2}}$. we then obtain the coefficient of the 
linear potential as  
\begin{equation}
  E=kr, \qquad  k=0.88{\rm \, GeV}/{\rm fermi}=0.176 ({\rm GeV})^2, 
\label{linearpot}
\end{equation}
which is identified with 
\begin{equation}
 k= \frac{1}{2\pi\alpha'}\,\, f(U_{\rm KK}) 
\end{equation}
in our model. 
Then, using the expression for $f(u)$ in Eq.~(\ref{QCD like model}),  we have 
\begin{equation}
 k= \frac{1}{2\pi\alpha'}\,\, f(U_{\rm KK})=\frac{1}{2\pi\alpha'}\,\,
\left(\frac{ U_{KK}}{R'}\right )^{3/2},
\end{equation}
where $R'$ and $U_{KK}$ are given in terms of $M_{KK}$ as
\begin{equation}
R'^{3}=\frac{2\,\pi\alpha_{c}N_{c}\, \alpha'}{M_{KK}},~~\mbox{and}~~ U_{KK}
=\frac{8\pi}{9}\,\, \alpha_{c}\, N_{c}\,\, \alpha'M_{KK}. 
\end{equation}
Therefore, we have the following expression of $k$ in terms of $M_{KK}$:
\begin{equation}
 k= \frac{8}{27}\,\,M_{KK}^2 \,\, (\alpha_c N_c).  
\end{equation}
We should remark here that in the above equation, 
the parameter $\alpha'$, the string tension of the order of the 
Planck scale in the original Lagrangian, disappears, and hence 
$k$ and $M_{KK}$ are the parameters of 
QCD scale.
By taking $ (\alpha_c N_c)\sim 1$,  we obtain 
\begin{equation}
 M_{KK} \approx \sqrt{\frac{27}{8}\times 0.176}\ \    {\rm GeV} \approx 0.77\,\, {\rm GeV}.  
\end{equation}

For convenience,  we express the parameters in the integration 
appearing in the above in terms of 
non-dimensional variables. The quark mass given in  Eq.~(\ref{quarkmass}) 
can be rewritten as  
\begin{equation}
m_i \equiv   \frac{U_{KK}}{2\pi\alpha'} \int^{v_i}_{1}\sqrt{f(v)g(v)}dv
= \frac{4M_{KK}}{9} \int^{v_i}_{1}\sqrt{\frac{1}{1-v^{-3}}}dv\, ,
\label{rquarkmass}
\end{equation}
with the non-dimensional parameter $v=\frac{u}{U_{KK}}$. 
The expression for the energy of a string given in Eq.~(\ref{totenergystring}) 
can be written as  
\begin{eqnarray}
E&=&\frac{1}{2\pi\alpha'}\, \left[ \int^{U_2}_{U_0} \pm \int^{U_1}_{U_0} \right] du \sqrt{ 
\frac{f(u)^3g(u)}{f(u)^2-f_0^2}} \\
&=&\frac{4M_{KK}}{9}\, \left[ \int_{v_0}^{v_2}\pm\int_{v_0}^{v_1}\right]
 v^3\sqrt{ \frac{1}{(v^3-v_0^3)(v^3-1)}}\, dv\, ,
\label{rtotenergystring}
\end{eqnarray}
with the length $r$ given by  
\begin{eqnarray}
r&=& f_0  \left[ \int^{U_2}_{U_0} \pm \int^{U_1}_{U_0} \right] du \sqrt{ 
\frac{g(u)}{f(u) \left(f(u)^2-f_0^2\right)}} \\
&=&\frac{3}{2M_{KK}} \left[ \int_{v_0}^{v_2}\pm\int_{v_0}^{v_1}\right] 
\sqrt{\frac{v_0^3}{(v^3-1)(v^3-v_0^3)}}\, dv\, ,
\label{rstringlength}
\end{eqnarray}
where $v_0=U_0/U_{KK}$.

The value $U_i=U_{KK}v_i$, determining the position of 
each flavor brane can be obtained from Eq.~(\ref{quarkmass}), 
once we fix the experimental values of 
the constituent quark masses $m_i$.
\footnote{The quark mass given by Eq. (\ref{rquarkmass}) is the constituent quark mass rather 
than the current quark mass, since
the non-perturbative QCD effects in this mass are 
 taken into account by the deformed space-time.} 
In Table \ref{ufrom}, we list the values of the 
constituent quark masses with the 
corresponding values of 
$U_i$ evaluated from  Eq.~(\ref{quarkmass}). 

Note that the parameters  $U_i$ are all expressed in units 
of $U_{KK}=\frac{8\pi}{9}\alpha'M_{KK} $,  with the  Planck scale 
parameter $\alpha'$ included. 
Therefore, the separation of the flavor branes 
in the extra direction is expressed in the extremely small units of 
$O({\rm GeV}/M_{\rm Planck}^2)=O(10^{-38} {\rm GeV}^{-1})$. 
However, this is the only parameter which explicitely includes 
the Plank scale parameter, $\alpha'$, 
 and all the other physical quantities,  such as the string 
energy $E$ and the separqation 
length $r$ in the Minkowski space,  are of the QCD scale ($\sim$  O(GeV)).

\begin{table}[htb]
%\begin{wraptable}{l}{\halftext}
\caption{The values  $U_i$ (in units of $U_{KK}$) which 
determine the positions of the flavor branes .}
\begin{center}
\begin{tabular}{|l|c|c|c|c|p{1cm}|} \hline\hline
  quark    &$  m_u$ &$m_s$   &$m_c$  &$m_b$ &$m_t$       \\ \hline
  mass(GeV) &0.363 & 0.546   & 1.5  &4.5 & 176          \\ \hline
$9m_i/4m_{KK}$  & 1.061 & 1.595& 4.383& 13.15&514.3 \\  \hline
$U_i$ (in the $U_{KK}$ unit) & 1.599 & 2.086& 4.826& 13.58&514.8 \\  \hline
\end{tabular}
\end{center}
%end{wraptable}
\label{ufrom}
\end{table}

\section{Numerical study of the meson string profile}
Having fixed the parameters in our QCD model, we next 
study how the shape of meson string is deformed when the separation distance, $r$, 
decreases. 

Let us consider a meson consisting of a light quark and a heavier 
anti-quark, which are located on
 the lower  and  higher branes, respectively. 
Here ``higher and lower'' means those of 
``larger and smaller'' values of $u$.  When the distance $r$ between a  quark and an  anti-quark 
becomes sufficiently small, the energetically favored shape 
is  such that the string starting from 
the lighter quark goes up directly to the heavier quark at 
the higher position, which differs from the standard 
shape, which goes down to a stationary point and then up to the heavier quark.  
In this way,  we can draw a rough sketch for the manner in which the  
shape of meson strings change depending on the separation distance $r$.

\begin{figure}
\begin{center}
\includegraphics[width=8cm,clip]{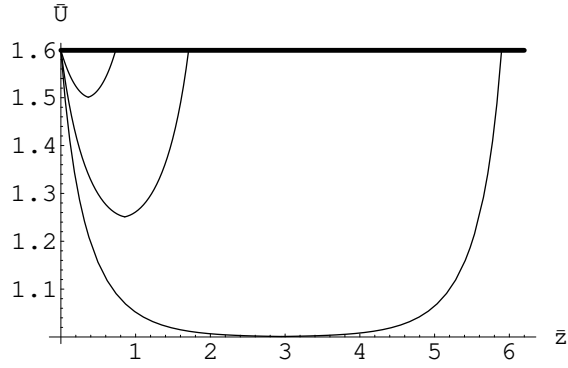}
\end{center}
\caption{Profiles of a $u\bar d$ string, where $\bar{z}=zM_{KK}$ and  $\bar{U}=u/U_{KK}$ 
 in the QCD model.}
\label{udstring}
\end{figure}

Now we report the results of a numerical calculation of the shape of meson strings, 
starting from $q_i$ of the $i$-th brane and 
stretching to $q_j$ on the $j$-th 
brane, by solving  Eq.~(\ref{eq:mesonshape}).
First, in  Fig.~{\ref{udstring}}, we display 
the shape of the $({u \bar d})$ string in the 
$(\bar{z},\bar{u})$ plane, where
\begin{equation}
\bar{z}=zM_{KK} ,
\end{equation}
\begin{equation}
\bar{U}=u/U_{KK} .
\end{equation}
It is seen that the shape changes with the 
the separation distance $r$,  as expected. 
If $r$ is very large, 
we see that the string almost  reaches to the minimal point,  
$U_{KK}$. This is a typical shape of a meson string.  
By contrast, in the 
small separation limit,  
it shrinks to zero length with a total energy tending to zero. 
\footnote{
We will come back to this short dintance behavior in a subsequent section. } 
%%%%%%%%%%%%%%%%%%%%%%%%%%%%%%%%%%%%
%%%%%%%%%%%%%%%%%%%%%%%%%%%%%%%%%%%%%%
%%%%%ud string %%%%%%%%%%%%%%%%%%%%%%%%%%%%%%%%%
%%%%%%%%%%%%%%%%%%%%%%%%%%%%%%%%%%%%%%
%%%%%%%%%%%%%%%%%%%%%%%%%%%%%%%%%%%%%%
%%%%%%%%%%%%%%%%%%%%%%%%%%%%%%%%%%%%
%%%%%%%%%%%%%%%%%%%%%%%%%%%%%%%%%%%%%%
%%%%%us string %%%%%%%%%%%%%%%%%%%%%%%%%%%%%%%%%
%%%%%%%%%%%%%%%%%%%%%%%%%%%%%%%%%%%%%%
%%%%%%%%%%%%%%%%%%%%%%%%%%%%%%%%%%%%%%
\begin{figure}
\begin{center}
\includegraphics[width=8cm,clip]{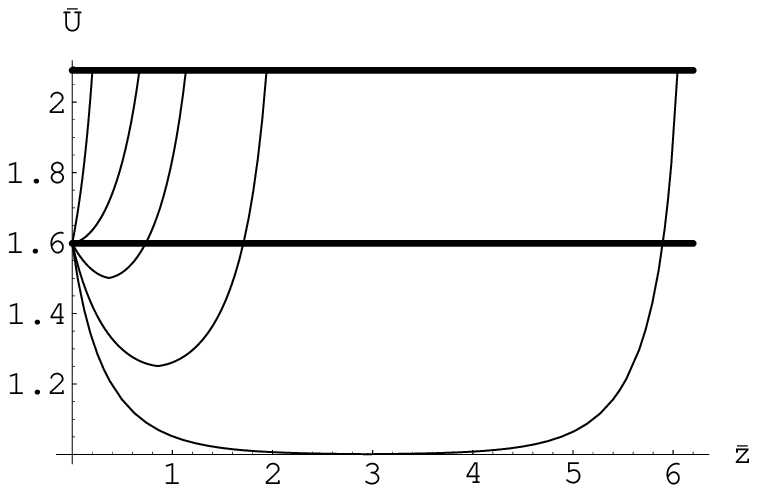}
\end{center}
\caption{Profiles of a $u\bar s$ string, where $\bar{z}=zM_{KK}$ and $\bar{U}=u/U_{KK}$
in the QCD-like model.}
\label{usstring}
\end{figure}
%%%%%%%%%%%%%%%%%%%%%%%%%%%%%%%%%%%%
%%%%%%%%%%%%%%%%%%%%%%%%%%%%%%%%%%%%%%
%%%%%ub string  %%%%%%%%%%%%%%%%%%%%%%%%%%%%%%%%%
%%%%%%%%%%%%%%%%%%%%%%%%%%%%%%%%%%%%%%
%%%%%%%%%%%%%%%%%%%%%%%%%%%%%%%%%%%%%%
\begin{figure}
\begin{center}
\includegraphics[width=8cm,clip]{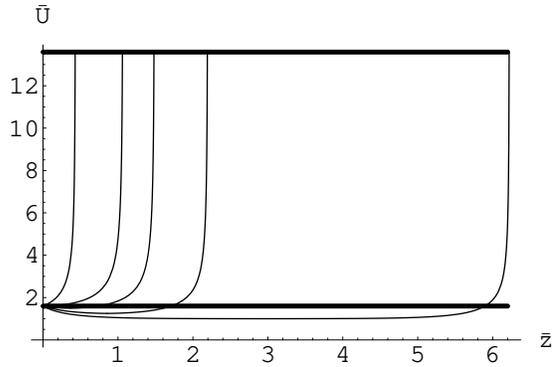}
\end{center}
\caption{Profiles of a $u\bar b$  string, where $\bar{z}=zM_{KK}$ and $\bar{U}=u/U_{KK}$ in the 
QCD like model.}
\label{ubstring}
\end{figure}%
Next, in the case of the  $({u \bar s})$ string, 
the shape is quite different from that of the  $({u \bar d})$ string. 
The calculated result for the string profile in the 
$(z,u)$ plane 
is shown in   Fig.~{\ref{usstring}}. We see that at 
large separations,  the string has almost the same 
shape as the $({u \bar d})$ string. 
However, if $r$ becomes smaller than the critical length 
$r_c(u\bar s)=0.9217\ {\rm GeV}^{-1} $ , it 
directly connects the two branes without passing through the 
region  $u \le U_1$. 
This tendency is more prominent in the shape of the $u \bar b$ string, 
as depicted in  Fig.~{\ref{ubstring}}. In this case,  the critical 
length is  $r_c(u\bar b)=1.427\  {\rm GeV}^{-1}$. 
%%%%%%%%%%%%%%%%%%%%%%%%%%%%%%%%%%%%
%%%%%%%%%%%%%%%%%%%%%%%%%%%%%%%%%%%%%%
%%%%%table of critical length %%%%%%%%%%%%%%%%%%%%%%%%%%%%%%%%%
%%%%%%%%%%%%%%%%%%%%%%%%%%%%%%%%%%%%%%
%%%%%%%%%%%%%%%%%%%%%%%%%%%%%%%%%%%%%%
\begin{table}[htb]
%\begin{wraptable}{l}{\halftext}
\caption{the values of $r_c$ for $u\bar d$,  $u\bar s$ and  $u\bar b$ strings. }
\label{ufromm}
\begin{center}
\begin{tabular}{|l|c|c|c|p{1cm}|} \hline\hline
string  &$u\bar d$    &$u\bar s$    &$u\bar c$ &$u\bar b$   \\ \hline
$r_c$ (in GeV$^{-1}$ )  &0 &0.9217& 1.352 &1.427       \\ \hline
$r_c M_{KK}$  &0 &0.710& 1.041 &1.099      \\ \hline
\end{tabular}
\end{center}
%\end{wraptable}
\end{table}
In order to derive physical meaning from our results, we 
study the potential between a quark and an anti-quark connected by a string in the next section.

\section{The potential between a quark and an antiquark}
The string potential at a finite distance $r=Z_2-Z_1$ is defined as 
\begin{equation}
V(r)= \frac{1}{2\pi\alpha'} \int^{U_2}_{U_1} \sqrt{ 
f(u) \left(f(u)dz^2+g(u)du^2\right)}-m_1-m_2.
\label{potential}  
\end{equation}
We have seen that the critical length $r_c$ becomes larger for strings 
connecting the 
$u$ quark of the first brane to the heavier anti-quark. 
Table 2 lists  the critical lengths for various mesons.
The existence of this critical length is a characteristic feature 
of a string connecting different flavors, and 
it is also very important to determine the quantitative features of the 
interaction between a quark and an anti-quark.

When we plot the potential  $V=V(r)$ between a $u$-quark and a $\bar{q_i}$-quark as a 
function of $r$, we need to connect the short distant part,  $r \le r_c$,  and the long distant 
part, $r \ge r_c$.
Thus, we can study the behavior of the potential by considering three regions for the distance, $r$:  
\begin{enumerate}
\item Region 1 : Long distance region ($\quad r >> r_c $) \\
From Eq.~(\ref{potential}),  it is easily seen that the potential 
converges to a single straight line, $ V(r)$. 
This iplies that, for the large separation case (region 1)  the potential is universal, 
and it is almost independent of the flavors of the quark and antiquark;  
\begin{equation}
V(r) \rightarrow  \  \frac{ f(U_{KK})}{2\pi\alpha'}\  r\,+C . 
\label{potrinf}  
\end{equation}
Here we have an almost $r$ independent constant term $C$, 
which comes from the difference between the real 
string energy and that calculated using an approximate in rectangular string, 
\begin{equation}
C \rightarrow_{r_c << r}E(r)-V({\rm basic\  rectangular}) \, ,  
\end{equation}
 where $C$ is almost independent of the quark and anti-quark flavors. 
This constant comes   from the
difference between the energy calculated from an exactly rectangular string 
around the point ($Z_2$, $U_{KK}$ ) and that of the 
real, smooth meson shape, 
which is almost independent of the flavors, because 
the difference exists only in the  region near the point   
 ($Z_2$, $U_{KK}$ ). 

\item  Region 2: Intermediate region ($\quad  r \sim r_c,$) \\
The critical distance $r_{c}$,  which separates  the short
  and long distant  parts of a potenntial, can be
 characterized as an inflection point for the potential $V(r)$,i.e.
\begin{equation}
\frac{d^2V(r)}{dr^2}\Biggr| _{r=r_{c}}=0.
\end{equation}
 The reason for this is as follows. 
From the above discussion, we understand that the derivative $u'(z)$ at the 
position of the u-quark endpoint in the short distance case and 
that in the long distance case have opposite signs, 
since a string in the former case is rising up from the position of the 
u-quark endpoint to that of the anti-quark endpoint, 
but in the latter case it goes down from the u-quark endpoint.
Therefore,  $u'(z)$ at the position of u-quark vanishes at the critical distance $r_{c}$.
From Eqs.~(\ref{eq:mesonshape}) and (\ref{mesonEnergy}),  we have 
\begin{equation}
\frac{dE}{dr}=\frac{dE}{du} u'
=\frac{1}{2\pi\alpha'}\, \frac{f^2}{f_0^2},
\end{equation}
from which we obtain
\begin{equation}
\frac{d^2E}{dr^2}=\frac{1}{2\pi\alpha'}\, \frac{2f}{f_0^2}\,
\frac{df}{du}\, u'
\end{equation}
Then, we can easily understand that the sign of $\frac{d^2E}{dr^2}$ is
determined by the sign of $u'(z)$. Therefore, $r_{c}$ becomes an inflection point 
for the potential $V(r)$, because $\frac{d^2E}{dr^2}=0$ at $r=r_{c}$, and 
is positive for $r \le r_{c}$ and negative
for $r \ge  r_{c}$. 
We have no such inflection point for mesons connecting branes of 
the same flavor.  The appearance of the inflection point is a 
characteristic feature of mesons connecting branes of different flavors. 
Note that the critical length depends on the flavor of the  meson;
in the case of the $u\bar d$ meson, $r_c=0$,  while $r_c=0.9217\, {\rm GeV}^{-1}$ 
for  $u\bar s$, and it becomes larger for mesons with heavier flavors. 
%%%%%%%%%%%%%%%%%%%%%%%%%%%%%%%%%%%%
%%%%%%%%%%%%%%%%%%%%%%%%%%%%%%%%%%%%%%
%%%%%potential for all the strings   %%%%%%%%%%%%%%%%%%%%%%%%%%%%%%%%%
%%%%%%%
%%%%%%%%%?=?????????????????????
%%%%%%%%%%%%%%%%%%%%%%
%%%%%%%%%%%%%%%%%%%%%%%%%%%%%%%%%%%%%%
\begin{figure}
\begin{center}
\includegraphics[width=15cm,clip]{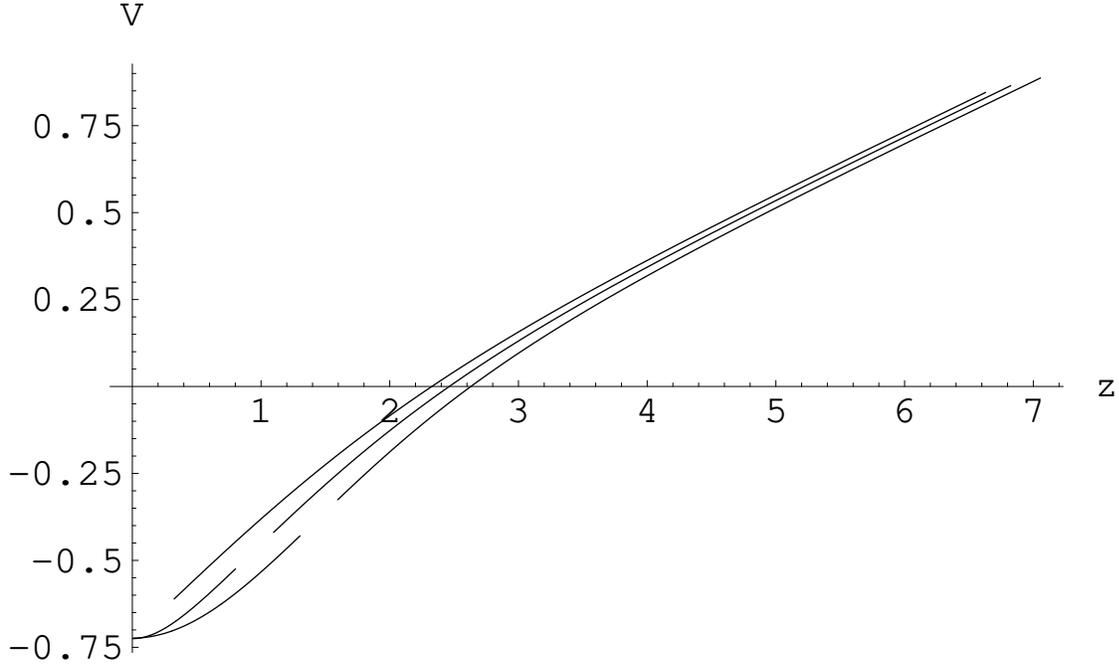}
\end{center}
\caption{The string potential as a function of $r=Z\  {\rm GeV}^{-1}$ for the strings  
$u\bar d$( the upper curve), $u\bar s$ ( the middle curve) and $u\bar b$ ( the lower curve).}
\label{potential}
\end{figure}%
%%%%%%%%%%%%%%%%%%%%%%%
%%%%%%%%%%%%%%%%%%%%%%%
\begin{figure}
\begin{center}
\includegraphics[width=15cm,clip]{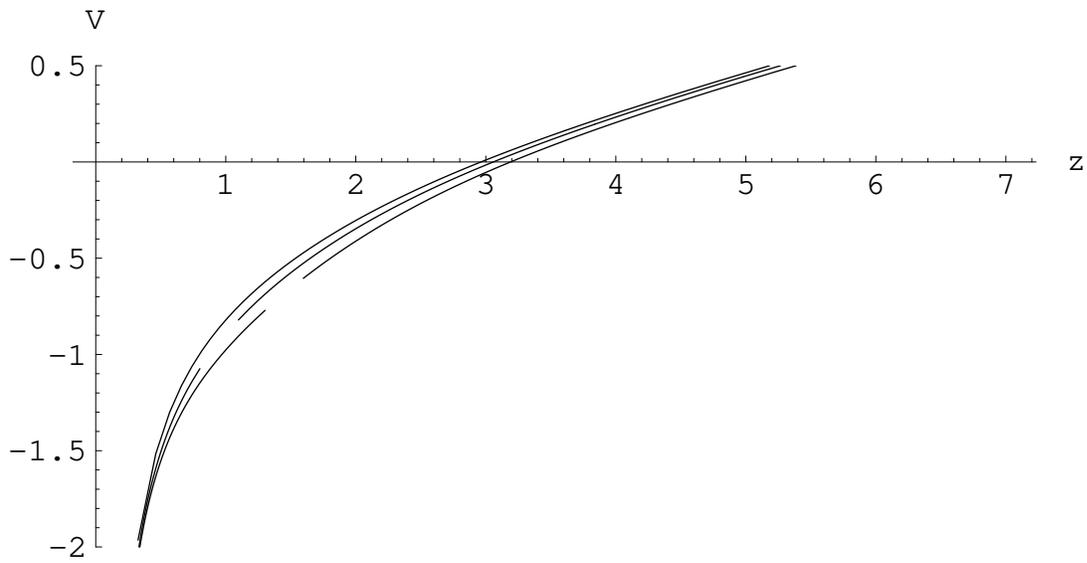}
\end{center}
\caption{The final form of the potential for a quark-antiquak 
pair obtained by taking account of the perturbative QCD effects added to the 
potential of the string connecting $q_{1}$ and $\bar q_{2}$ 
in the QCD-like model.}
\label{totpotential}
\end{figure}
\item Region 3: Short distance region ($\quad  r << r_c, $) \\
In the limit $r\rightarrow 0$, the string shape becomes 
almost  a straight line connecting the lighter quark $q_1$ to the heavier quark $q_2$, 
and there, the total energy of the meson becomes $m_2-m_1$.  
Subtracting the rest masses, the potential becomes 
\begin{equation}
V(r) \rightarrow   (m_2-m_1)-(m_1+m_2)=-2m_1,\quad m_1 \le m_2 \, .  
\label{potrzero}  
\end{equation}
\end{enumerate}
It is noted that in our string theory, eventhough it is called QCD-like, 
we cannot precisely take into account the short distance effects. Consequently, 
perturbative QCD effects resulting from one-gluon exchange would 
contribute as additional effects. 
 The dual gravity theory 
that  we study here  is effective in accounting for long distance non-perturbative 
effects of QCD, but it is ineffective in acconting for short distance perturbative effects 
of QCD. To implement the latter effects in the present framework, 
we have to estimate the higher-order effects. 
Here, we employ more convinient way to include such perturbative effects  
by taking account of the 
perturbative QCD results directly, 
as QCD is effective in the short distance region.  
The first order perturbative QCD potential is 
a sort of Coulomb potential, which is known to be expressed as  
\begin{equation}
V_{\rm perturbative\  QCD}=-\frac{4\, \alpha_{\rm QCD}}{3r}, \quad 
\alpha_{\rm QCD}=0.33, 
\label{pertqcdpot}
\end{equation}
which also is independent of the flavors of the quark and anti-quark. 

Therefore, a potential between the quark and anti-quark that is applicable to the short 
distance region as well as the long distance region may consist of the sum of 
the classical potential of the QCD-like string theory $V(r)$,  which we have studied 
in this paper,  and the one-gluon exchange potential $V_{\rm perturbative\  QCD}$ of 
QCD.  We call their  sum the \lq\lq improved potential": 
\begin{equation}
V(r)_{\rm improved} = V(r) + V_{\rm perturbative\  QCD}\, .
\label{improved}
\end{equation}
By including such perturbative effects of QCD, the short distance behavior of the potential 
$V(r)$ of Eq.(49) may be improved to some extent. 
Figure~\ref{totpotential} displays the calculated result for the improved potentials 
with different flavors.
\footnote{
If we take account of the spin dependent color-magnet interaction, they are, of course, flavor dependent.  Such an interaction modifies the potential only in the very short range region 
($\sim \delta(r)$), but we here consider the spin-independent potential by removing such spin-dependent 
effects.  
}

Although the behavior of the potential converges to a common value,  $-2m_1$, 
 in the limit $r\,\rightarrow \, 0$ (region 3), 
and also it converges to a single linear potential line in the large separation limit (region 1), 
the shape of the potential does depend on the flavors around the critical length, $r_c$ (region 2). 
It is interesting that the potential between the $u$ and $\bar q_j$ at relatively small 
distances becomes more attractive for the case as of $m_i$ increases.

From this picture, we can conclude that the flavor dependence of the potential can be 
found for the improved potentials around the critical region. 
It is noted that the non-perturbative effects can be included in our string picture 
realizing a linear potential which can be  directly derived from 
this QCD-like model. 

\section{Conclusion}
In the picture presented here, the origin of generation is attributed to the flavor branes. 
The flavor branes are separately positioned perpendicularly to the direction 
of the extra dimension $u$, which causes quite different energies 
for the meson strings, depending on the flavors.  

Based on the above picture, we have explored general formulations of 
the QCD-like string model and 
have obtained  classical solutions of meson strings. 
In this QCD-like model, it is automatical that 
the quark-antiquark potential is universal and independent of their flavors. 
This is consistent with the experimental data of  meson spectroscopy in the 
large separation limit. 
If the separation between a quark-antiquark pair decreases, however, 
their interaction becomes flavor dependent. 
As the most characteristic feature of our model, we have found that 
this attractive interaction becomes stronger  as the distance 
between the branes of the quark and  anti-quark increases 
depending on the difference of their flavors, especially in the middle 
region around critical 
separation,  $r \sim r_c$. 
If the endpoints of a string are on the same brane, there is 
no critical distance $r_c$. We believe  that our flavor dependent potential 
may be observed in flavored mesons whose endpoints lie on different branes.
In order to check this result, we need information concerning the   
quark-antiquark potential in the case that their endpoints lie on different flavored 
 branes. 
We hope that detailed analysis of hadron spectroscopy, especially 
for hadrons including different flavors, will clear 
whether or not our prediction is consistent with realistic QCD systems 
\cite{quiggrosner,cornellpot,stringmeson}.

The general formulae which we have used in this paper are 
also applicable to  more complicated shapes 
of hadron strings, 
because the differential equation 
Eq.~(\ref{eq:mesonshape}) is also applicable to various types of 
hadron strings, baryon strings, and more complicated web-like exotic hadronic strings 
possesing junctions. One 
 example  of such hadrons is 
the pentaquarks that we have investigated in Ref.\cite{BKST}. 
\footnote{Actually the observation of pentaquarks motivated us to formulate this picture of 
flavor branes.
It is found that experiments on multi-quark states have not yet been confirmed 
with the exception of $\Theta^{+}$,  which was first observed in Spring 8\cite{sp8}. 
Even though, we expect 
that quark-systems may have much variety of exotic hadrons 
to yield multi-quark systems 
including pentaquarks in addition to ordinary mesons and baryons.} 
In this respect, the string picture may suggest the new concept that 
``color and flavor are located at the 
endpoints of a string, 
but the spin, on the other hand, may be distributed over a whole string".  

If our interpretation of flavor in terms of flavor branes has 
some importance, our picture of the origin of flavor may give  new 
insight into hadron physics.  Then multi-quark states including quarks of 2nd or 3rd 
generations  may be packed in a more compact way; that is, 
if the distance between a quark and an anti-quark with different flavors 
becomes shorter than the critical length $r_c$, their interaction 
may become stronger, in which case, the density of the multi-quark system would be higher.  
Then, a variety of high density nuclear matters arise when different flavors are 
included. Such behavior  may be realized as a kind of crystal connecting quarks 
by junctions and  strings\cite{crystal}. 
This may be an indication of the existence of hadron states including strangeness with 
high density and it might be related to the recent observation of so-called 
 $\bar K$ nucleon clusters, which have high nucleon density 
at the center, 
specifically 
4-9 times higher than normal nuclear 
density ( $\rho =0.17\ fm^{-1}$)\cite{akaishiyamazaki}.  
Finally, our interpretation of flavors may not be irrelevant to the more 
speculative conjecture of the existence of a pasta phase of high density  
matter, appearing in the initial stage of supernova explosions\cite{sato}.  

Many problems are left to be solved. First, we ignored the effects of spin, and therefore, 
we cannot say definitely which meson, i.e., a pseudoscalar meson or a vector meson, 
corresponds to the string we 
have studied. 
Also the hyperfine interaction and parity should be taken into account in the string picture of hadrons. 
Recently,  studies of low energy QCD based on the AdS/CFT correspondence 
have been carried out many researchers; a holographic dual of QCD with massless 
flavors has been intensively studied by T.~Sakai and S.~Sugimoto\cite{SS}, who have elucidated 
the mechanism of chiral symmetry breakdown within this framework 
 and also derived the Chern-Simons term, which accouts for  
 baryon states as skyrmions. 
An interesting fact is that the model they studied is closely related to the hidden local symmetry 
approach\cite{BKM}. 
 Although the model beautifully expresses  
the mechanism of chiral symmetry breaking and the spin structure of the system, 
the  flavor structure with non-vanishing quark masses is 
 not well implemented. 
In this sense,  our model may be a complimentary approach. 
We hope a full understanding of non-perturbative QCD 
and the insight of the origin of flavor can be gained in near future. 

%--------------------------------
%
%         acknowledgements
%
%--------------------------------
\section*{Acknowledgements}
The authors are grateful to T.~Kugo for his kind and patient help  
and Y.~Imamura, S.~Sugimoto and T.~Sakai 
for useful discussions 
on string theories, especially on the AdS/CFT correspondence.  
Also,  we express our thanks to the partcipants at the international 
symposium  `` Flavor Physics and its Origin", 
 held at Ochanomizu University in December, 2005, for valuable comments.

M. B. and A. S. are partially supported by Grants-in-Aid for 
Scientific Research (No.17540238)  
from the Ministry of Education, Culture, Sports, Science and Technology, Japan.

%--------------------------------
%
%         bibliography
%
%--------------------------------

Note added: After submitting this paper we recieved  a letter informing us 
that the paper titled 
`` Multiflavor excited mesons from the fifth dimension".\cite{talavera} has  already 
proposed a methos to analyse meson strings very clearly, 
and it investigates  meson strings in detail 
on the basis of Mardacena metric, i.e., in the  SUSY case. 
The authors of that paper  focus mainly on spinning mesons and  intensively investigated
 Regge trajectories. 
Contrastly  we investigated  the quark-antiquark potential 
in the realistic case with SUSY breaking ( the 
Witten metric) in detail. 
We have found that the shape of potential differs on 
their flavors only in the middle region 
and that the shape of the potentials is almost independent of their flavors 
in the large separation limit. We found that the existence of the critical length is especially 
important. 
It would be intersting to compare their results in the SUSY case with 
our realistic results and to understand how 
the flavor branes play 
essential roles in flavor physics from QCD-like string models 
 in which 
the origin of generation comes from the existence of flavor branes.


\begin{thebibliography}{99}
\bibitem{BKST}
M.~Bando, T.~Kugo, A.~Sugamoto and Y.~Terunuma, Prog. Theor. Phys.. {\bf 2} (2004), 231; hep-ph/0410225.
\bibitem{S}
A. Sugamoto, talk given at 2nd international symposium on {\it ``New Developments of Integrated Sciences"} held at Ochanomizu U. on March 16 (2004);  hep-ph/0404019.
%\bibitem{KK}
%A. Karch and E. Katz, J.High Energy Phys. {\bf 06} (2002), 043; hep-th/0205236.

\bibitem{N=0a}
E. Witten, Adv. Theor. Math. Phys. {\bf 2} (1998), 505; hep-th/9803131;
J.High Energy Phys. 07 (1998), 006; hep-th/9805112.

D. J. Gross and H. Ooguri, Phys. Rev. {\bf D58} (1998), 106002; hep-th/9805129.

M. Kurczenski {\it et al.}, J.High Energy Phys. {\bf 05} (2004), 041; hep-th/0311270; 

\bibitem{M}
J. Maldacena, Adv. Theor. Math. Phys. {\bf 2} (1998), 231; hep-th/9711200.

J. Babington, J. Erdmenger, N. J. Evans, Z. Guralnik, and I. Kirsch Phys. Rev. {\bf D69} (2004), 066007; hep-th/0306018.

M. Kurczenski {\it et al.},J.High Energy Phys. {\bf 07} (2003), 049; hep-th/0304032.
\bibitem{Reggeslopehara}
S. Eidelman {\it et al.}, Phys. Lett. B {\bf 592}, (2004), 1; Review of particle properties.
\bibitem{quiggrosner}
C. Quigg and J.L. Rosner, Phys. Rev. {\bf D23} (1981), 2625.
\bibitem{cornellpot}
E. Eichten, K. Gottfried, T. Kinoshita, J. Kogut, K.D. Lane and T.M. Yan, 
Phys. Rev. Letters {\bf 34} (1975), 369.

%
%\bibitem{G}
%M. Gell-Mann, Phys. Lett. {\bf 8} (1964), 118.
%
\bibitem{stringmeson}
H. Boschi-Filho, Nelson R.F. Braga and C.N. Ferreira,  
hep-th/0512295.
\bibitem{sp8}
T. Nakano {\it et al.} (LEPS collaboration), Phys. Rev. Lett. {\bf 91} 
(2003), 012002.
\bibitem{akaishiyamazaki}
Y. Akaishi, A. Dote and T. Yamazaki, nucl-th/0501040v1 17 Jan 2005.
\bibitem{crystal}
Y. Igarashi, M, Imachi, T. Matsuoka, K. Ninomiya, S. Otsuki, S. Sawada and F. Toyoda, Suppliment of the Prog. THeor. Pysics, No. 63, 1978, P149.
\bibitem{sato}
G. Watanabe, T. Maruyama, K. Sato, K. Yasuoka and T. Ebisuzaki, 
Phys. Rev. Letters, 94, 031101(2005).

\bibitem{SS}
T. Sakai and S. Sugimoto, hep-th/0412141.
\bibitem{BKM}
M. Bando, T. Kugo and K. Yamawaki, Phys. Rep. {\bf 164} (1988), 217.
\bibitem{talavera}
Angel Paredes and Pere Talavera, Nucl.Phys.B713:438-464,2005
e-Print Archive: hep-th/0412260: also  see 
M. Kruczenski, L A. P. Zayas, J. Sonnenschein and D. Vaman, .JHEP 0506:046,2005, 
e-Print Archive: hep-th/0410035, for the flavor dependece 
of Regge trajectories for mesons in the holographic dual of large-N(c) QCD (although 
some parts, especially the part deriving the linear potential, include mistakes). 


\end{thebibliography}
\end{document}